\documentclass[usenatbib,referee]{mn2e}
\usepackage{graphicx}
\usepackage{epsfig}
\usepackage{color}

\title[The CD scenario for the progenitors of SNe Ia]
{The core-degenerate scenario for the progenitors of type Ia supernovae}
\author[B. Wang et al.]
{B. Wang,$^{\rm 1,2,3}$\thanks{E-mail:wangbo@ynao.ac.cn} W.-H. Zhou,$^{\rm 2,4}$ Z.-Y. Zuo,$^{\rm 5}$  Y.-B. Li,$^{\rm 6}$  
X. Luo,$^{\rm 7}$ J.-J. Zhang,$^{\rm 1,2}$ 
\newauthor  D.-D. Liu,$^{\rm 1,2,3}$ and C.-Y. Wu$^{\rm 1,2,3}$ \\
$^1$Yunnan Observatories, Chinese Academy of Sciences, Kunming 650216, China\\
$^2$Key Laboratory for the Structure and Evolution of Celestial Objects, Chinese Academy of Sciences, Kunming 650216, China\\
$^3$University of Chinese Academy of Sciences, Beijing 100049, China\\
$^4$Yunnan Minzu University, Kunming 650031, China\\
$^5$School of Science, Xi'an Jiaotong University, Xi'an 710049, China\\
$^6$Key Laboratory of Optical Astronomy, National Astronomical Observatories, Chinese Academy of Sciences, Beijing 100012, China\\
$^7$Department of Astronomy, Beijing Normal University, Beijing 100875, China}

\begin{document}
\date{Accepted. Received}
\pagerange{\pageref{firstpage}--\pageref{lastpage}} \pubyear{2016}
\maketitle
 
\label{firstpage}

\begin{abstract}
The origin of the progenitors of type Ia supernovae (SNe Ia) is
still uncertain. The core-degenerate (CD) scenario has been proposed as an alternative way for the production of SNe Ia.
In this scenario, SNe Ia are formed at the final stage of common-envelope evolution from a merger
of a carbon-oxygen white dwarf (CO WD) with the CO core of an asymptotic giant branch  companion.
However, the birthrates of SNe Ia from this scenario are still not well determined.
In this work, we performed a detailed investigation on the CD scenario based on a binary population synthesis approach. 
The SN Ia delay times from this scenario are  basically in the range of 90\,Myr$-$2500\,Myr,
mainly contributing to the observed  SNe Ia with short and intermediate delay times 
although this scenario can also produce some old SNe Ia.
Meanwhile, our work indicates that the Galactic birthrates of SNe Ia from this scenario are no more than 20\% of total SNe Ia
due to more careful treatment of mass transfer. 
Although the SN Ia birthrates in the present work are lower than those in Ilkov \& Soker, the CD  scenario
cannot be ruled out  as a viable mechanism for the formation of SNe Ia.
Especially, SNe Ia with circumstellar material from this scenario contribute to 0.7$-$10\% of total SNe Ia, 
which  means  that the CD scenario can reproduce the observed birthrates of SNe Ia like PTF 11kx.
We also found that SNe Ia happen systemically earlier for a high value of metallicity and their birthrates increase with metallicity.

\end{abstract}

\begin{keywords}
stars: evolution --  binaries: close  -- supernovae: general -- white dwarfs
\end{keywords}

\section{Introduction} 
Type Ia supernovae (SNe Ia) are among the most powerful explosions in the Universe 
and have high  scientific values in Cosmology, e.g. 
they are used to determine the cosmological parameters due to their high luminosities 
and remarkable uniformity (for a recent review see Howell 2011). 
They are also a predominant synthesis of chemical elements in their host galaxies, 
especially for the contribution of iron  (e.g. Matteucci \& Greggio 1986).
It has been accepted  that
SNe Ia arise from thermonuclear runaway explosions of carbon-oxygen white dwarfs (CO WDs)
in binaries, although their progenitor systems and
explosion mechanisms are still under debate (see, e.g. Podsiadlowski et al.\ 2008;
Hillebrandt et al. 2013; H\"{o}flich et al. 2013; Wang et al. 2013; Zhang et al. 2016).

Over the past four decades, two main families of SN Ia progenitor scenarios have been discussed frequently.
(1) \textit{The single-degenerate (SD) scenario} (e.g. Whelan \& Iben 1973; Nomoto, Thielemann \& Yokoi 1984).
In this scenario, a CO WD accretes hydrogen- or helium-rich  matter from a non-degenerate star to increase its mass close to the
Chandrasekhar limit and then results in a SN Ia explosion (e.g. Hachisu, Kato \& Nomoto 1996; Li \& van den Heuvel 1997; Han \& Podsiadlowski 2004; Meng, Chen \& Han 2009; Wang et al. 2009a;  L\"{u} et al. 2009; Chen \& Li 2009; Wu et al. 2016).
(2) \textit{The double-degenerate (DD) scenario}  (e.g. Webbink 1984; Iben \& Tutukov 1984). In this scenario,
a CO WD merges with another CO WD, the merging of which is due to the gravitational wave radiation that
drives orbital inspiral to merger, resulting in a SN Ia explosion (e.g. Nelemans et al. 2001; Geier et al. 2007; 
Chen et al. 2012; Ji et al. 2013; Ruiter et al. 2013; Liu et al. 2016).
Some variants of these two progenitor scenarios are needed to explain the observed diversity of SNe Ia
(for recent reviews see, e.g. Wang \& Han 2012; Maoz, Mannucci \& Nelemans 2014; Ruiz-Lapuente 2014).

Most simulations and calculations for the DD scenario relate to the merger 
of two cold CO WDs.\footnote{Note that the merger of two CO 
WDs may lead to an off-center carbon ignition, resulting in accretion induced collapse  
but not a SN Ia (e.g. Saio \& Nomoto 1985, 2004; Timmes, Woosley \& Taam 1994).}
Meanwhile, a CO WD can also merge with the hot CO core of an asymptotic giant branch (AGB) star, and then
produce a SN Ia, which is known as the core-degenerate (CD) scenario 
(e.g. Kashi \& Soker 2011; Ilkov \& Soker 2012; Soker 2013).
The merger of a WD with the hot core of an AGB star
was first investigated by Sparks \& Stecher (1974)
who suggested that a SN could be formed through this type of merger.
Livio \& Riess (2003) argued that the merger of
the WD with the AGB core can explain the presence of hydrogen lines in some observed SNe
Ia such as SN 2002ic when the merger happens at the end of the common-envelope (CE) stage or shortly after.
Importantly, the CD scenario may  avoid the off-center carbon ignition 
during the merger process  (e.g. Soker 2013).

In the CD scenario, a SN Ia explosion may occur shortly or a long time after the CE stage (e.g. Soker et al. 2013).
Kashi \& Soker (2011) suggested that this scenario could form massive WDs 
with super-Chandrasekhar mass, leading to the formation of super-luminous SNe Ia (see also Ilkov \& Soker 2012). In addition,
a violent  prompt merger via the CD scenario may  reproduce the properties of 
some SNe Ia with circumstellar material (CSM) such as PTF 11kx (e.g. Soker et al. 2013). 
This scenario was also used to explain the properties of SN 2011fe and SN 2014J (e.g. Soker et al. 2014; Soker 2015).
Aznar-Sigu\'{a}n et al. (2015) recently performed three-dimensional smoothed particle hydrodynamics simulations
of the merger stage between the WD and the AGB core, and argued that this merge process can result in the formation of a massive
WD and then produce a SN Ia.  We note that
Tsebrenko \& Soker (2015) recently summarized the properties of the CD scenario and
made detailed comparisons between this scenario and other progenitor scenarios.

Although the CD scenario may explain some properties of SN Ia diversity,
SN Ia birthrates from this scenario are still not well determined 
(e.g. Ilkov \& Soker 2013; Aznar-Sigu\'{a}n et al. 2015; Zhou et al. 2015).
Ilkov \& Soker (2013) argued that this  scenario can reproduce
the observed birthrates of total SNe Ia based on a simplified binary population synthesis (BPS) code.
Tsebrenko \& Soker (2015) recently  estimated that more than 20\% of all SNe Ia come from the CD
scenario on the basis of the fraction of SNe Ia that happen inside planetary nebulae.
Briggs et al. (2015) also carried out some BPS studies for the merger of WD$+$AGB core,
and argued that the majority of  magnetic WDs with strong fields are the carbon-oxygen type and may merge
within a CE. Thus, Briggs et al. (2015) suggested that the merger of WD$+$AGB core may be relatively common.

The purpose of this article is to investigate SN Ia birthrates and delay times for the CD scenario
using a detailed Monte Carlo BPS approach.
We describe the numerical methods and assumptions for the BPS approach in Sect. 2, and give the
BPS results in Sect. 3. Finally, a discussion and summary are presented in Sect. 4.

\section{Numerical Methods }

\subsection{Physical input}
In the CD scenario, a Chandrasekhar or super-Chandrasekhar mass WD
could be formed through the merger of a cold CO WD with the hot CO core of an AGB star.
A series of Monte Carlo BPS simulations for the CD scenario are performed in this study. 
We adopted the following assumptions as the criteria for producing  SNe Ia 
through the CD scenario (e.g. Soker 2013; Ilkov \& Soker 2013):
(1)  The combined mass of the  CO WD  (${M}_{\rm WD}$, primary) and the AGB core (${M}_{\rm core}$, secondary) 
during the final stage of CE evolution is larger than or equal to the Chandrasekhar mass limit
(1.378$M_{\odot}$ in this work), that is, ${M}_{\rm WD}+{M}_{\rm core} \geq 1.378M_{\odot}$.\footnote{  
Nomoto, Thielemann \& Yokoi (1984) suggested that  the critical mass limit of a cold non-rotating WD for carbon 
ignition is about $1.378\,M_{\odot}$.  Note that the Chandrasekhar mass limit for WDs is 1.44$M_{\odot}$ in
Newtonian gravity, which drops to 1.4$M_{\odot}$ when general relativity
is taken into account (see Mathew \& Nandy 2014).}
(2) The WD and the AGB core merge during the final stage of CE evolution, in which the mass of  the AGB core (${M}_{\rm core}$)
is  limited to be lower than $1.1\,M_{\odot}$ to avoid the formation of ONe cores that cannot produce SNe Ia.

The CO WD is usually disrupted and
accreted onto the more massive AGB core during the merging process. 
However, in some conditions the AGB core would be disrupted and accreted onto the cooler CO
WD if ${M}_{\rm WD}$ is larger than ${M}_{\rm core}$ (e.g. Soker et al. 2013). 
In such case, the  SN explosion may occur  shortly after the CE stage, resulting in
a SN Ia explosion inside a planetary nebula shell, which may reproduce the 
properties of some SNe Ia with CSM such as PTF 11kx. This case is 
known as the violent prompt merger  scenario (e.g. Soker et al. 2013), 
which is studied  in the present work.
Additionally, we also examined the influence of metallicity on the birthrates of SNe Ia for the CD scenario, in which
metallicities were chosen to be Z$=$0.0001, 0.004, 0.02 and 0.03.

The delay times of SNe Ia
are defined as the timescale from the formation of primordial  binaries to SN explosions. In the CD scenario,
the theoretical delay times of SNe Ia  are the sum of the evolutionary timescale 
from primordial binaries to the formation of CD systems and the spin-down 
timescale from the merger product of WD$+$AGB core to SN explosion.  
The spin-down timescale of the merger product is mainly determined by 
the magneto-dipole radiation torque, which
can be written as
\begin{equation}
\tau_{\rm B}\approx5\times10^{\rm
7}\left(\frac{B\sin\delta}{10^{\rm 7} {\rm G}}\right)^{\rm -2}
{\rm yr},
\end{equation}
where $B\sin\delta$ follows a distribution of
 \begin{equation}
 \frac{dN}{d\log(B\sin\delta)}={\rm constant},
 \end{equation}
for $10^{\rm 6} {\rm G}\leq B\sin\delta \leq 10^{\rm 8} {\rm G}$
(e.g. Ilkov \& Soker 2012; Meng et al. 2012). According to this
distribution, we can get the spin-down timescale using a Monte
Carlo simulations.  Note that
there is no time delay by spin-down or gravitational wave radiation for the case of these
violent prompt mergers (e.g. Soker et al. 2013), and the spin-down
timescale can be neglected for those with merging masses above
1.48$M_{\odot}$ in our simulations.  The massive WDs over 1.48$M_{\odot}$ 
require differential rotation for support before explosions (e. g. Yoon \& Langer 2004).
The spin-down time of these WDs  is mainly determined by the 
time-scale of internal angular-momentum redistribution, 
which may be extremely short as the angular momentum transport by the 
Eddington-Sweet meridional circulation is relatively fast  (see, e.g.  Yoon \& Langer 2004; Saio \& Nomoto 2004; 
Piro 2008; Hachisu et al. 2012).

\subsection{Basic parameters for Monte Carlo simulations}

We carried out a series of Monte Carlo BPS simulations for the CD scenario. 
In each simulation, we followed the evolution of $1\times10^{\rm 7}$ primordial binaries
from star formation to the formation of WD$+$AGB systems 
using the Hurley binary evolution code (see Hurley, Tout \& Pols 2002).
We simply supposed a constant star formation rate (SFR) of  $5\,{M}_{\odot}\rm yr^{-1}$
over the past 14\,Gyr or, alternatively, it is
supposed as a delta function, i.e. a single instantaneous starburst. 
The constant SFR is used to  approximate  spiral galaxies,
whereas the delta function provides a rough description of globular clusters or elliptical
galaxies.

The Monte Carlo BPS simulations require as input
the initial mass function (IMF) of the primordial stars, 
the  initial mass-ratio distribution, 
the distribution of initial orbital separations, 
and the eccentricity distribution of binary orbit.
The Monte Carlo BPS studies are highly dependent on the chosen initial conditions.
The following initial parameters for the Monte Carlo simulations are adopted:
 (1) The initial mass function (IMF) for the primordial primary is taken from Miller \& Scalo (1979, MS79).
 Alternatively, we also consider the IMF of Scalo (1986, S86).
 (2)  A constant mass-ratio distribution is supposed (e.g. Mazeh et al. 1992; Goldberg \& Mazeh 1994).
Alternatively, we also adopt an uncorrelated mass-ratio distribution, in which both binary components 
are chosen independently from the same IMF.
 (3) The distribution of initial orbital separations is supposed to be constant in $\log a$ for
wide binaries, in which $a$ is the orbital separation 
(for more discussions see Eggleton, Fitchett \& Tout  1989; see also Han et al. 1995).
 (4) All stars are supposed to be members of binaries.
The primordial binaries are generated through a Monte Carlo method, and
a circular orbit is supposed for all binaries. It is usually not necessary to model an initial eccentricity  distribution of binary orbit in
BPS studies as binary systems tend to circularise before interacting (see Hurley, Tout \& Pols 2002). 

\subsection{Evolutionary way to WD$+$AGB systems}
SN Ia explosions in the CD scenario originate from  the evolution of WD$+$AGB systems. 
In Fig. 1, we present the binary evolutionary way to form WD$+$AGB systems. 
The primordial binary system in the CD scenario has a wide separation, which allows the primordial primary to 
evolve into the AGB stage. The primordial primary first fills its Roche lobe when it evolves to AGB stage 
(it now contains a CO core in its center).
Before the primordial primary fills its Roche-lobe, it loses a lot of 
matter through the stellar wind in the AGB stage, 
which leads to the subsequent stable Roche-lobe overflow (RLOF).
In the Hurley binary evolution code, the treatment of RLOF is a revised version originated from Tout et al. (1997),
and the radius-mass exponent $\zeta$ defined by Webbink (1985) is employed to deal with the stability of mass transfer.
At the end of RLOF, the  primary  become  a CO WD. 
Meanwhile, the MS secondary  become a massive star due to the stable mass transfer, resulting in the formation of a CO WD$+$MS system.
The CO WD$+$MS system continues to evolve, and the MS secondary may fill its Roche-lobe again when it
evolves to the AGB stage. At this stage, a CE may be formed due to the dynamically unstable mass transfer, resulting from
the deep convective envelope of the AGB star and the large mass-ratio. 
The CO WD will merge with the CO core of the AGB star during the CE stage if the 
CE cannot be ejected. Finally, a SN Ia is produced at the final stage of the CE evolution
(see also Ilkov \& Soker 2013).
 
For the CD scenario,  SN Ia explosions for the ranges of
the initial mass of the primordial primary are from $2.0$-$6.5\,M_{\odot}$, 
and the initial mass of the primordial  secondary  from 1.5$-$6.0$\,M_{\odot}$. 
The initial  orbital period of the primordial system needs to be wide enough  
(e.g. larger than 8\,yr) so that the primordial primary can evolve to an AGB star before it fills its Roche-lobe.

\begin{figure}
\begin{center}
\epsfig{file=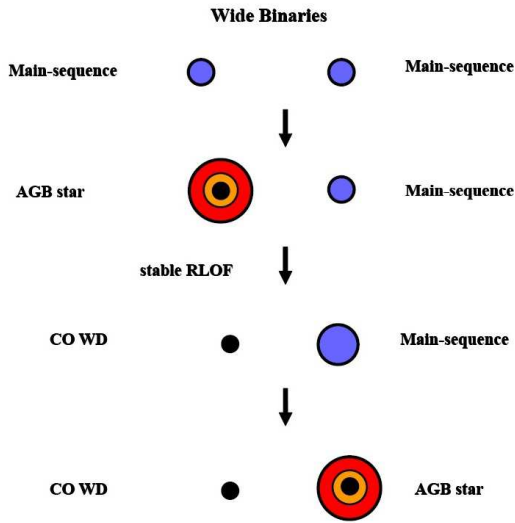, width=7.5cm} \caption{Binary evolutionary way to WD$+$AGB systems.}
\end{center}
\end{figure}

\subsection{Common envelope evolution}

In the CD scenario, the merger of WD$+$AGB core originates from the  CE evolution.  
In oder to obtain the output of the CE stage,
the standard energy equation is adopted
(e.g. Webbink 1984). The CE  is supposed to be ejected once
\begin{equation}
 \alpha_{\rm CE} \left( {G M_{\rm don}^{\rm f} M_{\rm acc} \over 2 a_{\rm f}}
- {G M_{\rm don}^{\rm i} M_{\rm acc} \over 2 a_{\rm i}} \right) = {G
M_{\rm don}^{\rm i} M_{\rm env} \over \lambda R_{\rm don}},
\end{equation}
where  $G$ is the gravitational constant,
$\alpha_{\rm CE}$ is the CE ejection efficiency, 
$M_{\rm acc}$ is the mass of the accreting star, $M_{\rm don}$ is the mass of the
donor, $a$ is the orbital separation, $R_{\rm don}$ is the  radius of the donor,
$M_{\rm env}$ is the envelope mass of the donor, 
$\lambda$ is a structure parameter of the donor  relevant to the stellar mass-density distribution, 
and the indices ${\rm i}$ and ${\rm f}$ indicate the initial and final values during the CE stage.
The left side of this equation presents  the difference of orbital energy between the final and initial stage,
whereas the right side  gives the binding energy of the CE.

There are two highly uncertain parameters in equation (3), that is,   $\alpha_{\rm CE}$ and $\lambda$,
which are key parts of our understanding the  orbital evolution of  the binary  during the CE stage.
Previous studies usually combined $\alpha_{\rm CE}$ and $\lambda$ into a single free parameter $\alpha_{\rm CE}\lambda$
when calculating the output of CE stage (e.g. Wang et al. 2009b).
The value of $\lambda$ depends on the structure and evolutionary stage of the
donor, which is usually set to be
0.5 for simplicity (e.g. Hurley et al. 2002). However, the real value of $\lambda$ 
may be far from 0.5 (e.g. van der Sluys, Verbunt \& Pols 2006; see also Dewi \& Tauris 2000).
In this paper, we used the fitting formulae of
envelope binding energy $E_{\rm bind}$ of the original giant,
which implicitly contain the variable $\lambda$ (see Loveridge, van der Sluys \& Kalogera 2011). 
The $E_{\rm bind}$ is estimated
by integrating the gravitational and internal energies from the core-envelope
boundary (${M_{\rm c}}$) to the surface of the star ($M_{\rm s}$),
\begin{equation}
E_{\rm bind}=\int^{M_{\rm s}}_{M_{\rm c}}(-\frac{Gm}{r(m)})dm+\alpha_{\rm in}\int^{M_{\rm s}}_{M_{\rm c}}E_{\rm in}dm,
\end{equation}
in which $\alpha_{\rm in}$ is the percentage of the internal energy contributing to
the ejection of the envelope and $E_{\rm in}$ is the internal energy including the thermal energy of
the gas and the radiation energy (e.g. van der Sluys, Verbunt \& Pols 2006; Zuo \& Li 2014). 
$\alpha_{\rm in}$ is usually set to be 1 though its value is rather uncertain.
Here,  we take  $\alpha_{\rm in}$ as 1 in our basic model  and change  it to zero  for comparisons (see set 3 in Table 1).
Meanwhile, we also change the values  of $\alpha_{\rm CE}$ (e.g. 0.1, 0.2, 0.3 and 0.5)  to examine its influence on the final results.

\section{Results}
\subsection{Birthrates of SNe Ia}

\begin{table}
 \begin{center}
 \caption{Galactic SN Ia birthrates for different  Monte Carlo BPS simulation sets. 
 Notes: 
 $\alpha_{\rm CE}$ = CE ejection efficiency; 
 $\alpha_{\rm in}$ =  percentage of the internal energy contributing to the ejection of the envelope; 
 ${\rm IMF}$ = initial mass function;
$n(q)$ = initial mass-ratio distribution; 
$\nu_{\rm CD}$\,($\nu_{\rm CSM}$)= SN Ia birthrates for the CD  scenario (SNe Ia with CSM); 
$\nu_{\rm SD}$ = SN Ia birthrates for the SD scenario that includes the contributions of 
WD$+$MS/RG channels based on the studies of Wang, Li \& Han (2010);
$\nu_{\rm DD}$ = SN Ia birthrates for the DD scenario based on the studies of Wang et al. (2010).}
   \begin{tabular}{cccccccc}
\hline \hline
Set & $\alpha_{\rm CE}$ & $\alpha_{\rm in}$ & ${\rm IMF}$ & $n(q)$ & $\nu_{\rm CD}$\,($\nu_{\rm CSM}$) 
& $\nu_{\rm SD}$ & $\nu_{\rm DD}$\\
&&&&&($10^{-3}$\,yr$^{-1}$)&($10^{-3}$\,yr$^{-1}$)&($10^{-3}$\,yr$^{-1}$)\\
\hline
$1$ & $0.1$   & $1$  & ${\rm MS79}$   & ${\rm Constant}$  & $0.852\,(0.566)$      & $1.768$ & $0.707$\\
$2$ & $0.2$   & $1$  & ${\rm MS79}$   & ${\rm Constant}$  & $0.440\,(0.323)$      & $1.480$  & $0.158$\\
$3$ & $0.2$   & $0$  & ${\rm MS79}$   & ${\rm Constant}$  & $0.479\,(0.295)$      & $0.899$ & $0.358$\\
$4$ & $0.2$   & $1$  & ${\rm S86}$      & ${\rm Constant}$  & $0.373\,(0.279)$      & $1.072$ & $0.133$\\
$5$ & $0.2$   & $1$  & ${\rm MS79}$   & ${\rm Uncorrelated}$  & $0.034\,(0.029)$      & $0.257$ & $0.011$\\
$6$ & $0.3$   & $1$  & ${\rm MS79}$   & ${\rm Constant}$   & $0.116\,(0.090)$     & $1.221$ & $0.077$\\
$7$ & $0.5$   & $1$  & ${\rm MS79}$   & ${\rm Constant}$   & $0.050\,(0.041)$     & $1.329$ & $0.100$\\
\hline
\end{tabular}
\end{center}
\end{table}

The observed SN Ia  birthrate  in our Galaxy  is about 4$\times 10^{-3}\rm yr^{-1}$, which
can be used to constrain the progenitor models of SNe Ia (e.g. Cappellaro \& Turatto 1997). 
In order to systematically study the Galactic SN Ia birthrates
for the CD scenario, we carried out seven sets of Monte Carlo BPS simulations with metallicity
$Z=0.02$  (see Table 1).  In these simulations,
we varied the initial input parameters to examine their influences on the BPS results.
For comparisons, we also show the results of SNe Ia for the SD and DD scenarios in this table. 
We found that the birthrates of SNe Ia are sensitive to uncertainties in some model
parameters based on the seven sets of simulations.
For example, if we adopt a mass-ratio distribution with uncorrelated binary components,
the birthrates of SNe Ia  will decrease significantly (set 5).

\begin{figure}
\begin{center}
\epsfig{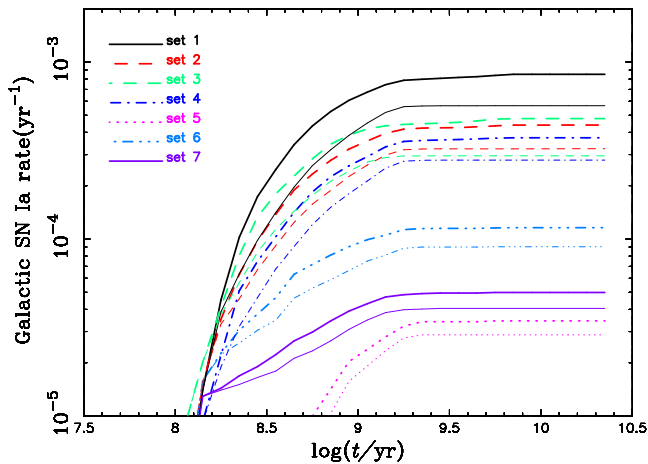} \caption{Evolution of the Galactic SN 
Ia birthrates with time for a  constant Population I SFR  with different BPS simulation sets. The thick lines
are for all SNe Ia from the CD scenario, whereas the thin lines 
are only for the SNe Ia with CSM like PTF 11kx. }
\end{center}
\end{figure}

In Fig. 2, we show the evolution of
the Galactic SN Ia birthrates with time for the CD scenario by adopting metallicity $Z=0.02$ and 
a  constant SFR of $5\,M_{\odot} \rm yr^{-1}$. According to  the seven sets of Monte Carlo simulations, 
the theoretical  birthrates from this scenario are 
in the range of 0.5$-$8.5$\times10^{-4}\rm yr^{-1}$,
accounting for 1$-$20\% of the observations.  
The SN Ia birthrates from the CD scenario are lower than those in observations, which means that
the CD scenario can only form part of total SNe Ia.
We note that the birthrates of SNe Ia from the CD scenario fall with increasing 
the value of  $\alpha_{\rm CE}$.
This is because  a high value of  $\alpha_{\rm CE}$ leads to wider orbital separation, 
resulting in few mergers of WD$+$AGB core.
If the CD scenario can really contribute to a high proportion of observed 
SNe Ia, a low value of $\alpha_{\rm CE}$ is expected.

Soker et al. (2013) recently suggested that the violent prompt merger in the CD scenario
may account for the very massive  CSM  in PTF 11kx, and for the presence of hydrogen shells in the CSM.
According to our BPS simulations, the  Galactic birthrates for SNe Ia with CSM  are in 
the range of 0.3$-$5.7$\times10^{-4}\rm yr^{-1}$, accounting for 0.7$-$10\% of total SNe Ia,
which can reproduce the  birthrates of SNe Ia like PTF 11kx;  
the observed fraction of  SNe Ia with CSM is estimated  to be 0.1$-$1\% (e.g. Dilday et al. 2012).

\subsection{Delay times of SNe Ia}

\begin{figure}
\begin{center}
\epsfig{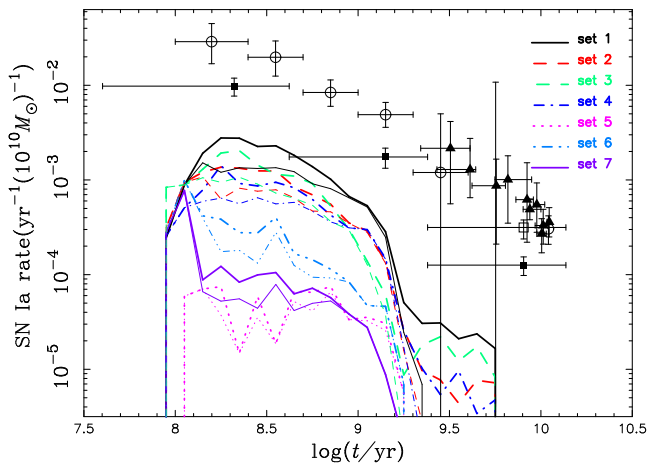} \caption{Delay-time distributions 
of SNe Ia with different BPS simulation sets, where 
the spin-down time is included in the delay time.  The thick lines
are for all SNe Ia from the CD scenario, whereas the thin lines are only for SNe Ia with CSM like PTF 11kx.
The open circles are taken from Totani et al. (2008), the open square is from Graur \& Maoz (2013), 
the filled triangles and squares are  from Maoz, Keren \& Avishay  (2010) and 
Maoz, Mannucci \& Timothy (2012), respectively.}
\end{center}
\end{figure}

The delay time distributions of SNe Ia can be obtained from observations, which 
could be used to constrain the progenitor models of SNe Ia. 
In Fig. 3, we present the SN Ia delay time distributions for the CD scenario 
based on a single starburst of $10^{\rm 10}M_{\odot}$,
where the spin-down time is included in the delay time. 
The estimated delay times for this scenario are  mainly in the range of 
90\,Myr$-$2500\,Myr after the starburst, which may have a 
contribution to the SNe Ia with short and intermediate delay times. 
For the SNe Ia with CSM, the SN explosion may occur shortly after the CE stage.
The delay time of SNe Ia in this case is mainly determined by the evolutionary timescale 
from primordial binaries to the formation of CD systems. 
Note that the shortest delay times of SNe Ia for  the CD scenario are mainly 
determined by the MS lifetime of a star with mass about $7\,M_{\odot}$.

As expected in some previous studies (e.g. Kashi \& Soker  2011; Ilkov \& Soker 2012), 
we also found that the CD scenario can even form some  SNe Ia 
with delay times as long as several Gyr for a low value of  $\alpha_{\rm CE}$  (e.g. $\alpha_{\rm CE}=0.1$ and 0.2);
a low value of  $\alpha_{\rm CE}$ tends to form more mergers with masses below
1.48$M_{\odot}$, and thus a long spin-down time.
The delay time of SNe Ia in this case is mainly determined by 
the spin-down time after the merging of the WD+AGB core. 
If the spin-down time is too long (e.g. several Gyr),   the mergers may experience
post-crystallization stage,
leading to the formation of a highly carbon enriched outer layer. 
Thus,  Soker et al. (2013) suggested that this case may
account for  the carbon-rich composition of the fastest-moving ejecta of SN 2011fe. 

\subsection{Mass distribution of WD$+$AGB core }

\begin{figure}
\begin{center}
\epsfig{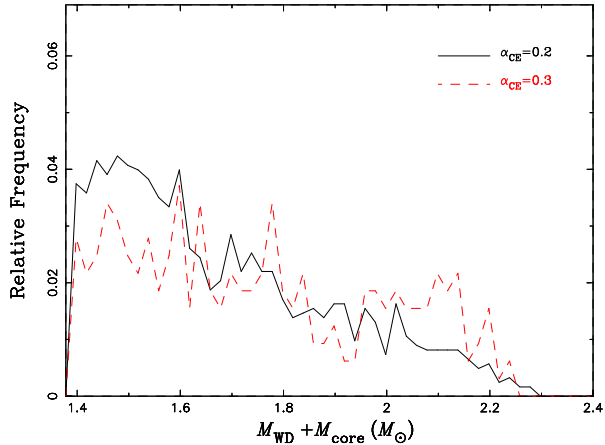} \caption{Distribution of the total mass of 
${M}_{\rm WD}$  and ${M}_{\rm core}$ in  the WD$+$AGB 
systems that can ultimately form SNe Ia. The black solid and red dashed lines 
present the cases of $\alpha_{\rm CE}= 0.2$  (set 2)  and $\alpha_{\rm CE}= 0.3$  (set 6) , respectively.
The number in every case is normalized to 1.}
\end{center}
\end{figure}

\begin{figure}
\begin{center}
\epsfig{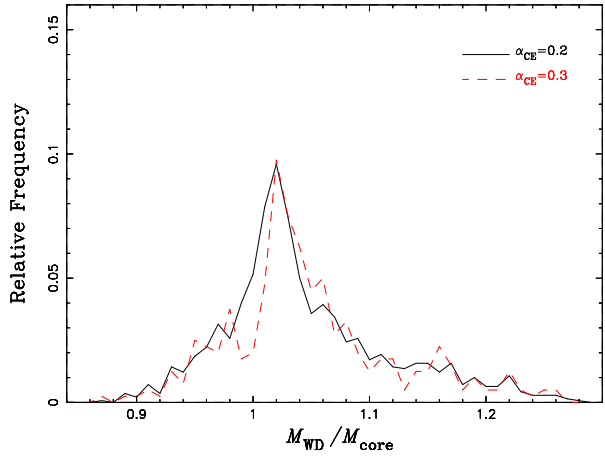} \caption{Same as Fig. 4, but for the distribution of 
mass ratio between ${M}_{\rm WD}$ and ${M}_{\rm core}$. }
\end{center}
\end{figure}

\begin{figure}
\begin{center}
\epsfig{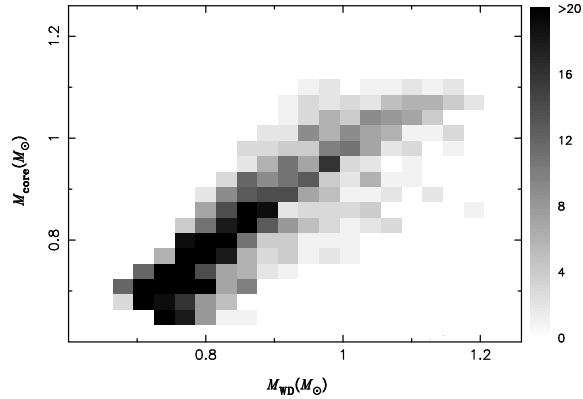} \caption{Distribution of ${M}_{\rm WD}$ and 
${M}_{\rm core}$ in WD$+$AGB systems that can ultimately form SNe Ia, where $\alpha_{\rm CE} = 0.2$  (set 2).}
\end{center}
\end{figure}

The WD explosion mass may have an influence on the final production of  nickel-56 and thus the maximum luminosity 
of SNe Ia  (e.g. Arnett 1982).  
In Fig. 4, we show the distribution of the combined masses of ${M}_{\rm WD}$ 
and ${M}_{\rm core}$ in the WD$+$AGB  systems that can ultimately form SNe Ia through the CD scenario. The combined masses 
have a peak nearby  $1.4{M}_{\odot}$ and then decrease, which 
can be understood by the initial mass function of stars. 
These massive systems with combined masses  above 2$M_{\odot}$ may contribute to over-luminous SNe Ia
that have been observed with inferred WD explosion masses in the order of 2$M_{\odot}$ 
(e.g. Howell et al. 2006; Hicken et al. 2007; Scalzo et al. 2010).  
Note that the SD scenario of SNe Ia can also form over-luminous SNe Ia 
if the rotation of WDs is considered, in which the rotating WDs have 
been prevented from exploding until angular-momentum redistribution 
 (e.g. Justham 2011; Hachisu et al. 2012; Wang et al. 2014).

Fig. 5 presents the distribution of mass ratio between ${M}_{\rm WD}$ and ${M}_{\rm core}$ 
with different values of $\alpha_{\rm CE}$.  
From this figure, we can see that almost all the values of ${M}_{\rm WD}/{M}_{\rm core}$ are above 0.9, and
there is a peak nearby 1.0.
In Fig. 6, we show the distribution of ${M}_{\rm WD}$ and ${M}_{\rm core}$ in the WD$+$AGB systems that can ultimately form SNe Ia. 
In this figure,  WD$+$AGB systems with  ${M}_{\rm core} > 1.1 M_{\odot}$ are rejected
as more massive CO cores in AGB stars have already burnt all their central carbon  and will form ONe cores  (e.g. Umeda et al. 1999).

\subsection{The  effect of metallicity}

\begin{figure}
\begin{center}
\epsfig{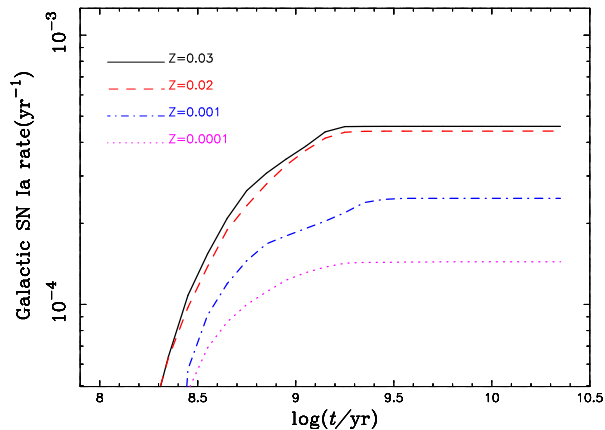}
\caption{The evolution of SN Ia birthrates for a constant SFR with different metallicities of $Z=0.03$, 
0.02, 0.001 and 0.0001, where $\alpha_{\rm CE} = 0.2$.}
\end{center}
\end{figure}

Metallicities may have important effects on the properties of SNe Ia
(e.g. Timmes, Brown \& Truran 2003; Podsiadlowski et al. 2008; Sullivan et al. 2010).
In Fig. 7, we present the evolution of SN Ia birthrates for a constant SFR of $5\,M_{\odot} \rm yr^{-1}$ with different metallicities.
From this figure, we can see that metallicity has a significant influence on the birthrates of SNe Ia, especially for low metallicity environments;
the birthrates increase with metallicity and SNe Ia happen systemically earlier for a high value of metallicity. 
The WD$+$AGB systems mainly originate from massive primordial binaries that tend to exist in high 
metallicity environments (e.g. Wang \& Han 2010).
Thus, SN Ia birthrates from the CD scenario increase with metallicity. 
Meanwhile, massive primordial binaries evolve more quickly than low-mass ones,
resulting in a systematically earlier SN explosion for a high value of metallicity.

\section{Discussion and Summary}

\begin{figure}
\begin{center}
\epsfig{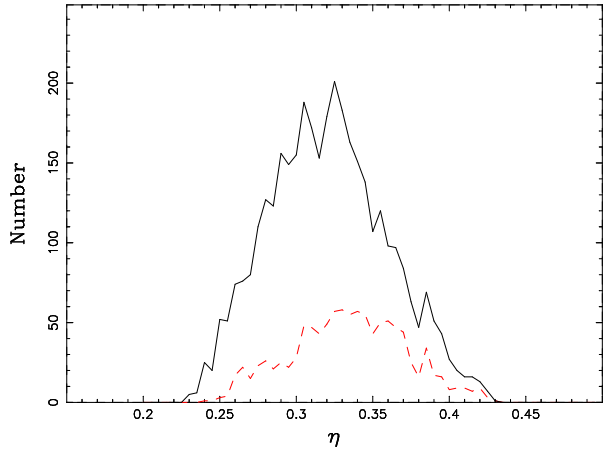}
\caption{Distribution of the mass-transfer parameter $\eta$ in our simulations, where $\alpha_{\rm CE} = 0.2$  (set 2). 
The black solid  line presents the distribution of $\eta$ for all potential WD$+$AGB systems, 
whereas the red dashed line shows the distribution of $\eta$ for these WD$+$AGB systems that can lead to SNe Ia.}
\end{center}
\end{figure}

According to a simple BPS code, 
Ilkov \& Soker (2013)  suggested that  the CD scenario can reproduce the observed birthrates of SNe Ia,
and claimed that this  scenario plays an important role for producing SNe Ia. 
However,  we found that this scenario can only account for  part  (1$-$20\%) of SNe Ia based on our detailed BPS approach. 
The main difference between  these two works is the treatment of the binary interaction during mass transfer.
Ilkov \& Soker (2013)  obtained 
the new mass of the primordial secondary after the primordial primary passed through the AGB stage and
became a WD (with mass ${M}_{\rm WD}$) as
${M}_{\rm 2new}={M}_{2}+\eta ({M}_{1}-{M}_{\rm WD})$, where $\eta$ is the mass-transfer parameter,  
${M}_{1}$ and ${M}_{2}$ are the initial masses of the primordial primary and secondary, respectively.
In Fig. 8, we show the distribution of  the mass-transfer parameter $\eta$ in our simulations.
From this figure, we can see that the value of  $\eta$  in our simulations is  mainly in the range of 0.25$-$0.4 
that is much lower than the value ($\eta$$\sim$0.8$-$0.9) taken 
by Ilkov \& Soker (2013). The high mass-transfer parameter  in Ilkov \& Soker (2013) leads to the formation of  
more massive AGB stars and larger CO cores, 
and thus gives a much higher birthrates of SNe Ia.

In this work, we assume that SNe Ia in the CD scenario  can be 
produced during the  final stage of  CE evolution. However,
a SN Ia  can also be formed through the CD scenario when the merging process of double WDs happens 
shortly after  the CE  stage (e.g. within about $10^{5}$\,yr), the merging of which 
is due to the gravitational wave radiation (e.g. Soker 2013). 
Meng \& Yang (2012) recently estimated that  SN Ia birthrates from this case is very low ($<$0.1\% to all SNe Ia). 
Thus, we speculate that the condition studied here dominates the production of SNe Ia in the  CD scenario 
although this scenario only contributes to part of the observed SNe Ia. 

Soker et al. (2013) recently suggested that the violent prompt merger in 
the CD scenario may explain some SNe Ia with very massive CSM such as PTF 11kx.
If we adopted a strict assumption on the total mass of the WD$+$AGB core and the envelope mass (${M}_{\rm env}$) of the AGB star
(e.g. ${M}_{\rm WD}+{M}_{\rm core} \geq 1.8 M_{\odot}$ and ${M}_{\rm env} \geq 0.5 M_{\odot}$; Soker et al. 2013),  
the SN Ia birthrates from the violent prompt merger scenario will decrease to 
0.3$-$8.0$\times10^{-5}\rm yr^{-1}$,  
accounting for 0.1$-$2\% of all SNe Ia, 
which can still reproduce the observational number  of SNe  Ia like PTF 11kx. 

By employing  a detailed BPS approach,
we got an upper limit for the birthrates of SNe Ia based on the CD scenario (no more than 20\% of total SNe Ia). 
The birthrates in our simulations are lower than those in Ilkov \& Soker (2013), 
the main reason of which is that we adopted a detailed mass-transfer process.
We found that the CD scenario mainly contributes to the SNe Ia with short and intermediate delay times
although this scenario can also produce some SNe Ia with long delay times.
The birthrates of SNe Ia with CSM are estimated to be 0.7$-$10\% of total SNe Ia, 
which can match the observed number of SNe Ia  like PTF 11kx.
We also found that  SNe Ia happen systemically earlier for a high value 
of metallicity and their birthrates increase with metallicity.
In order to put further constraints on the CD scenario, more numerical 
simulations and observational evidence for this scenario are 
needed. 
 
\section*{Acknowledgments}
We acknowledge the  referee, Christopher Tout, for the valuable comments that helped us
to improve the paper.  
This work is supported by the National Basic Research Program of China (973 programme, 2014CB845700),
the National Natural Science Foundation of China (Nos 11673059, 11322327, 11390374, 11521303, 
11303036, 11573021, 11403096 and 61561053),  
the Chinese Academy of Sciences (Nos QYZDB-SSW-SYS001, XDB09010202 and KJZD-EW-M06-01), 
the Fundamental Research Funds for the Central Universities,
the Natural Science Basic Research Program of Shaanxi Province - Youth Talent Project (No. 2016JQ1016),
the Natural Science Foundation of Yunnan Province (Nos 2013HB097 and 2013FB083),  
and  the Youth Innovation Promotion Association CAS.

\label{lastpage}

\begin{thebibliography}{}
\bibitem[Arnett (1982)]{Arn82}                             Arnett W. D., 1982, ApJ, 253, 785
\bibitem[Aznar-Sigu\'{a}n et al. (2015)]{Azn15}  Aznar-Sigu\'{a}n G., Garc\'{\i}a-Berro E., Lor\'{e}n-Aguilar P., Soker N., Kashi A., 2015, MNRAS, 450, 2948
\bibitem[Briggs et al. (2015)]{Bri15}                     Briggs G. P. et al., 2015, MNRAS, 447, 1713
\bibitem[Cappelllaro (1997)]{Cap97}                     Cappellaro E., Turatto M., 1997, In: Ruiz-Lapuente P., Cannal R., Isern J. (Eds.), Proceedings of the 
NATO Advanced Study Institute, Thermonuclear Supernovae, Vol. 486. Kluwer, Dordrecht, p. 77
\bibitem[Chen \& Li (2009)]{che09}                     Chen W.-C., Li X.-D., 2009, ApJ, 702, 686
\bibitem[Chen et al. (2012)]{che12}                      Chen X., Jeffery C. S., Zhang X., Han Z., 2012, ApJ, 755, L9
\bibitem[Dewi \& Tauris (2000)]{dew00}              Dewi J. D. M., Tauris T. M., 2000, A\&A, 360, 1043
\bibitem[Dilday et al.  (2012)]{dil120}                  Dilday B. et al., 2012, Science, 337, 942
\bibitem[Eggleton, Fitchett \& Tout  (1989)]{eft89}  Eggleton P. P., Fitchett M. J., Tout C. A.,  1989, ApJ, 347, 998
\bibitem[Geier et al. (2007)]{Ger07}                     Geier S., Nesslinger S., Heber U., Przybilla N., Napiwotzki R., Kudritzki R. P., 2007, A\&A, 464, 299
\bibitem[Graur \& Maoz (2013)]{Gra13}              Graur O.,  Maoz D., 2013, MNRAS, 430, 1746
\bibitem[Hachisu, Kato \& Nomoto (1996)]{hac96}   Hachisu I., Kato M., Nomoto K., 1996, ApJL, 470, L97
\bibitem[Hachisu et al. (2012)]{hac12}                  Hachisu I., Kato M., Saio H., Nomoto K., 2012, ApJ, 744, 69
\bibitem[Han \& Podsiadlowski (2004)]{han04}   Han Z., Podsiadlowski P., 2004, MNRAS, 350, 1301
\bibitem[Han et al. (1995)]{Han95}                       Han Z., Podsiadlowski P.,  Eggleton P. P., 1995, MNRAS, 272, 800
\bibitem[Hicken (2007)]{Hic07}                            Hicken M. et al., 2007, ApJ, 669, L17
\bibitem[Hillebrandt et al. (2013)]{hill13}             Hillebrandt W., Kromer M., R\"{o}pke F. K., Ruiter A. J., 2013, FrPhy, 8, 116
\bibitem[H\"{o}flich et al. (2013)]{hoe13}            H\"{o}flich P. et al., 2013, FrPhy, 8, 144
\bibitem[Howell (2011)]{How11}                          Howell D. A., 2011, Nature Communications, 2, 350
\bibitem[Howell (2006)]{How06}                          Howell D. A. et al., 2006, Nature, 443, 308
\bibitem[Hurley, Tout \& Pols (2002)]{Hur02}      Hurley J. R., Tout C. A., Pols O. R., 2002, MNRAS, 329, 897
\bibitem[Iben \& Tutukov (1984)]{IT84}               Iben I., Tutukov A. V., 1984, ApJS, 54, 335
\bibitem[Ilkov (2012)]{Ilk12}                                Ilkov M., Soker N., 2012, MNRAS, 419, 1695
\bibitem[Ilkov (2013)]{Ilk13}                                Ilkov M., Soker N., 2013, MNRAS, 428, 579
\bibitem[Ji et al. (2013)]{ji13}                               Ji S. et al., 2013, ApJ, 773, 136
\bibitem[Justham (2011)]{jus11}                            Justham S., 2011, ApJ, 730, L34
\bibitem[Kashi (2011)]{Kas11}                              Kashi A.,  Soker N., 2011, MNRAS, 417, 1466
\bibitem[Li \& van den Heuvel (1997)]{lix97}      Li X.-D.,  van den Heuvel E. P. J., 1997, A\&A, 322, L9
\bibitem[Liu et al. (2016)]{Liu16}                         Liu D.,Wang B., Podsiadlowski P., Han Z., 2016, MNRAS, 461, 3653
\bibitem[Livio (2003)]{Liv03}                               Livio M., Riess A., 2003, ApJ, 594, L93
\bibitem[Loveridge, van der Sluys \& Kalogera (2011)]{Lov11}   Loveridge A. J., van der Sluys M. V., Kalogera V., ApJ, 2011, 743, 49
\bibitem[L\"{u} et al. (2009)]{lv09}                      L\"{u} G., Zhu C., Wang Z., Wang N., 2009, MNRAS, 396, 1086
\bibitem[Maoz, Mannucci \& Nelemans (2014)]{mao14}   Maoz D., Mannucci F., Nelemans G., 2014, ARA\&A, 52, 107
\bibitem[Maoz, Keren \& Avishay  (2010)]{mao10}            Maoz D., Keren S., Avishay G.-Y., 2010, ApJ, 722, 1879
\bibitem[Maoz, Mannucci \& Timothy (2012)]{mao12}      Maoz D., Mannucci F., Timothy D. B., 2012, MNRAS, 426, 3282
\bibitem[Mathew \& Nandy  (2014)]{mao14}      Mathew A.,  Nandy  M. K., 2014, 	preprint (arXiv:1401.0819)
\bibitem[Matteucci \& Greggio (1986)]{Mat86}  Matteucci F., Greggio L., 1986, A\&A, 154, 279
\bibitem[Meng, Chen \& Han (2009)]{men09}     Meng X., Chen X., Han Z., 2009, MNRAS, 395, 2103
\bibitem[Meng \& Yang (2012)]{Men12}             Meng X., Yang W., 2012, A\&A, 543, A137
\bibitem[Miller \& Scalo (1979)]{Mil79}             Miller G. E., Scalo J. M., 1979, ApJS, 41, 513 (MS79)
\bibitem[Nelemans et al. (2001)]{nel01}              Nelemans G., Yungelson L. R., Portegies Zwart S. F., Verbunt F., 2001, A\&A, 365, 491
\bibitem[Nomoto, Thielemann \& Yokoi  (1984)]{Nom84}     Nomoto K., Thielemann F.-K.,  Yokoi K., 1984, ApJ, 286, 644
\bibitem[Piro (2008)]{Pir08}                                Piro A. L., 2008, ApJ, 679, 616
\bibitem[Podsiadlowski (2008)]{Pod08}              Podsiadlowski P., Mazzali P., Lesaffre P., Han Z.,  F\"orster F., 2008, New Astron. Rev., 52, 381
\bibitem[Ruiter et al. (2013)]{Rui13}                   Ruiter A. J. et al., 2013, MNRAS, 429, 1425
\bibitem[Ruiz-Lapuente (2014)]{rui14}               Ruiz-Lapuente P., 2014, New Astron. Rev., 62, 15
\bibitem[Saio \& Nomoto (1985)]{sn85}             Saio H., Nomoto K., 1985, A\&A, 150, L21
\bibitem[Saio \& Nomoto (2004)]{sn04}             Saio H., Nomoto K., 2004, ApJ, 615, 444
\bibitem[Scalo (1986)]{Sca86}                            Scalo J. M., 1986, Fund. Cosm. Phys., 11, 1 (S86)
\bibitem[Scalzo (2010)]{Sca10}                          Scalzo R. A. et al., 2010, ApJ, 713, 1073
\bibitem[Soker (2013)]{Sok13}                           Soker N., 2013, IAUS, 281, 72
\bibitem[Soker (2015)]{Sok15}                           Soker N., 2015, MNRAS, 450, 1333
\bibitem[Soker et al. (2013)]{SK13}                   Soker N. et al., 2013, MNRAS, 431, 1541
\bibitem[Soker et al. (2014)]{SG14}                   Soker N. et al., 2014, MNRAS, 437, L66
\bibitem[Sparks (1974)]{Spa74}                         Sparks W. M.,  Stecher T. P., 1974, ApJ, 188, 149
\bibitem[Sullivan et al. (2010)]{sul10}               Sullivan M. et al., 2010, MNRAS, 406, 782
\bibitem[Timmes, Woosley \& Taam (1994)]{TIM94} Timmes F. X., Woosley S. E., Taam R. E., 1994, ApJ, 420, 348
\bibitem[Timmes, Brown \& Truran (2003)]{TIM03}  Timmes F. X., Brown E. F., Truran J. W., 2003, ApJ, 590, L83
\bibitem[Totani (2008)]{Tot08}                          Totani T., Morokuma T., Oda T., Doi M., Yasuda N., 2008, PASJ, 60, 1327
\bibitem[Tout et al. (1997)]{Tou97}                    Tout C. A., Aarseth S. J., Pols O. R., Eggleton P. P., 1997, MNRAS, 291, 732
\bibitem[Tsebrenko \& Soker (2015)]{tse15}      Tsebrenko D., Soker N., 2015, MNRAS, 447, 2568
\bibitem[Umeda et al. (1999)]{ume99}                Umeda H., Nomoto K., Yamaoka H., Wanajo S., 1999, ApJ, 513, 861
\bibitem[van der Sluys, Verbunt \& Pols (2006)]{vand06}  van der Sluys M. V., Verbunt F., Pols O. R., 2006, A\&A, 460, 209
\bibitem[Wang et al. (2009a)]{wan09a}                   Wang B., Meng X., Chen X., Han Z., 2009a, MNRAS, 395, 847
\bibitem[Wang et al. (2009b)]{wan09b}                   Wang B., Chen X., Meng X., Han Z., 2009b, ApJ, 701, 1540
\bibitem[Wang et al. (2010)]{wan10}                   Wang B. et al., 2010, Sci. China Ser. G, 53, 586
\bibitem[Wang \& Han (2010)]{wh10}                 Wang B., Han Z., 2010, A\&A, 515, A88
\bibitem[Wang \& Han (2012)]{wh12}                 Wang B., Han Z., 2012, New Astron. Rev., 56, 122
\bibitem[Wang et al. (2014)]{wan14}                    Wang B., Justham S., Liu Z., Zhang J., Liu D., Han Z., 2014, MNRAS, 445, 2340
\bibitem[Wang, Li \& Han (2010)]{wan10}           Wang B., Li X.-D., Han Z., 2010, MNRAS, 401, 2729
\bibitem[Wang et al. (2013)]{wxf13}                    Wang X.-F., Wang L., Filippenko A. V., Zhang T., Zhao X., 2013, Science, 340, 170
\bibitem[Webbink (1984)]{Web84}                      Webbink R., 1984, ApJ, 277, 355
\bibitem[Webbink (1985)]{Web85}                      Webbink R., 1985, in Interacting Binary Stars, eds. J. E. Pringle, \& R. A. Wade (Cambridge University Press), 39
\bibitem[Whelan (1973)]{Whe73}                        Whelan J.,  Iben I., 1973, ApJ, 186, 1007
\bibitem[Wu et al. (2016)]{Wu16}                       Wu C., Liu D., Zhou W., Wang B., 2016, RAA, 16, 160
\bibitem[Yoon \& Langer (2004)]{yoo04}           Yoon S.-C., Langer N., 2004, A\&A, 419, 623
\bibitem[Zhang et al. (2016)]{zhan16}                 Zhang J.-J. et al., 2016, ApJ, 817, 114
\bibitem[Zhou et al. (2015)]{zhou15}                  Zhou W., Wang B., Meng X., Liu D., Zhao G., 2015, RAA, 15, 1701
\bibitem[Zuo \& Li (2014)]{zuo14}                     Zuo Z.-Y., Li X.-D., 2014, MNRAS, 442, 1980

\end{thebibliography}
\end{document}